\documentstyle[12pt,epsfig]{article}
\begin{document}
\renewcommand{\thefootnote}{\fnsymbol{footnote}}
\sloppy
\newcommand{\rp}{\right)}
\newcommand{\lp}{\left(}
\newcommand \be  {\begin{equation}}
\newcommand \bea {\begin{eqnarray}}
\newcommand \ee  {\end{equation}}
\newcommand \eea {\end{eqnarray}}

\title{The Nasdaq crash of April 2000: \\
Yet another example of log-periodicity in a speculative
bubble ending in a crash}

\author{Anders Johansen$^1$ and Didier Sornette$^{1,2,3}$\\
$^1$ Institute of Geophysics and
Planetary Physics 3845 Slichter Hall, Box 951567\\
University of California, Los Angeles, California 90095-1567\\
$^2$ Department of Earth and Space Science\\
University of California, Los Angeles, California 90095\\
$^3$ Laboratoire de Physique de la Mati\`{e}re Condens\'{e}e\\ CNRS UMR6622 and
Universit\'{e} de Nice-Sophia Antipolis\\ B.P. 71, Parc
Valrose, 06108 Nice Cedex 2, France}

\date{3 May 2000}

\maketitle

\abstract{The Nasdaq Composite fell another $\approx 10 \%$ on Friday the
14'th of April
2000 signaling the end of a remarkable speculative high-tech bubble
starting in
spring 1997. The closing of the Nasdaq Composite at $3321$ corresponds to a
total loss of
over $35 \%$ since its all-time high of $5133$ on the 10'th of March 2000.
Similarities to the speculative bubble preceding the infamous crash of
October 1929 are quite striking: the belief in what was coined a ``New
Economy''
both in 1929 and presently made share-prices of companies with three digits
price-earning ratios soar. Furthermore, we show that the largest draw downs
of the Nasdaq are outliers with a confidence level better than $99\%$ and that
these two speculative bubbles,
as well as others, both nicely fit into the quantitative framework proposed
by the authors in a series of recent papers.}

\maketitle

\thispagestyle{empty}
\pagenumbering{arabic}
\newpage

\section{Introduction}

A series of recent papers \cite{SJB,SJ97,risk,JSL,JLS,Emerg} have presented
increasing evidence that market crashes as well as large corrections are
often preceded by speculative bubbles with two main characteristics:
a power law acceleration of the market price decorated with log-periodic
oscillations. Here, ``log-periodic'' refers to the fact that the oscillations
are periodic in the logarithm of the time-to-crash. Specifically, it has been
demonstrated that the equation
\bea \label{lpeq}
F\lp t\rp = A + B\lp t_c -t\rp^z + C\lp t_c -t\rp^z\cos\lp \omega \ln \lp
t_c - t\rp - \phi\rp
\eea
remarkably well quantifies the time-evolution of the bubble in terms of the
price ending with a
crash or large correction at a time close to $t_c$. This equation corresponds
to a first order Fourier expansion of the general power law solution to a
``renormalization equation''
\bea
\frac{dF\lp t\rp}{d\ln\lp t_c -t\rp} = z + i\omega
\eea
around the ``critical point'' $t_c$. Quite remarkable, for all the bubbles in
the most liquid markets, {\it e.g.}, U.S.A., Hong-Kong and the Foreign
Exchange
Market, the log-frequency $\omega/2\pi$ have consistently been close to $1$.
Within the framework of power laws with complex exponents, or equivalently
discrete scale invariance \cite{sordsi}, this corresponds to a prefered
scaling
ratio $\lambda \approx e \approx 2.7$: the local period of the log-periodic
oscillations decreases according to a geometrical series with the ratio
$\lambda$. For a range of emergent markets, larger fluctuations were seen in
the value of $\lambda$, but the statistics resulting from over twenty bubbles
were quite consistent with that of the larger markets \cite{Emerg}. In
contrast, the ``universality'' of the value of the real part of the exponent
quantifying the acceleration in the price has not been established. From a
theoretical view point, this is not surprising: a rational expectation model
of bubbles and crashes show
that depending on whether the size of the crash is proportional to the price
itself or that of the increase due to the bubble, either the logarithm of the
price or the price itself is the correct quantity characterising the bubble
\cite{risk}. Another more technical reason for the larger fluctuations in the
exponent $z$ comes from the well-known sensitivity in the determination of
critical exponents due to finite-size-effects as well as errors in the
determination of the value of the critical point $t_c$.

The question is whether we, in an objective and non-arbitrary manner, can
define
a crash and hence when we should expect eq. (\ref{lpeq}) to be a good
description of the preceding bubble? This is the subject of the next section.

\section{Crashes are Outliers} \label{out}

It is well-known that the distributions of stock market returns exhibit
``fat tails''. For example, a $5\%$ daily loss in the Dow Jones Industrial
Average occurs approximately once every two years while the Gaussian framework
would predict one such loss in about a thousand year. Furthermore, the
unconditional volatility on various emergent markets is much higher than on
developed equity markets \cite{Bekaert}. These empirical observations has
led to the development of more sophisticated models than the Gaussian,
for instance involving power law tails
\cite{Koedijk,Vries,Pagan,Olsen,stanley} or stretched
exponentials \cite{lahsor} as well as models allowing for non-stationary
of volatility such
as ARCH and GARCH models \cite{GARCH}, which better reproduces the
statistics of the
market fluctuations. Crashes on the other side are the most extreme events
and there are two possibilities to describe them\,:

\begin{enumerate}
\item The distribution of returns is stationary and the extreme events can be
extrapolated as lying in its far tail. Within this point of view, recent works
in finance and insurance have recently investigated the relevance of the body
of theory known as Extreme Value Theory to extreme events and crashes
\cite{Embrechts1,Embrechts2,Embrechts3}.

\item Crashes cannot be accounted for by an extrapolation of the distribution
of smaller events to the regime of extremes and belong intrinsically to another
regime, another distribution, and are thus outliers.

\end{enumerate}

In order to see which one of these two descriptions is the most
accurate, a statistical analysis of market fluctuations \cite{JS98.1} was
performed. Instead of looking at the usual ``one-point statistics'' in terms
of the distribution of returns, higher order correlations were included by
instead considering so-called {\it draw downs}. A draw down is defined as a
persistent decrease in the index over consecutive days. Specifically, the
daily closing of the Dow Jones was considered disregarding occasional single
upwards movements of less than 1\%. It was established that the
distribution of draw downs
of the Dow Jones Average daily closing from 1900 to 1993 is well approximated
by an exponential distribution with a decay constant of about $2\%$. (As we
will soon see the decay is actually slower than that of an exponential). This
exponential distribution holds only for draw downs smaller than about $15\%$.
In other words, this means that all draw downs of amplitudes of up to
approximately $15\%$ are well approximated by the same exponential distribution
with characteristic scale $2\%$. This characteristic decay constant means that
the probability of observing a draw down larger than $2\%$ is about $37\%$.
Following hypothesis 1 and extrapolating this description to, {\it e.g.},
the three largest crashes on the USA market in this century (1914, 1929 and
1987) yields a recurrence time of about $50$ centuries for {\it each single}
crash. In reality, the three crashes occurred in less than one century.

As an additional null-hypothesis, $10.000$ synthetic data sets, each covering
a time-span close to a century hence adding up to about $10^6$ years, have been
generated using a GARCH(1,1) model estimated from the true index with a
t-student distribution with four degrees of freedom \cite{GARCH}. This model
includes both non-stationarity of volatilities and fat tail nature of the price
returns. In conclusion, our analysis \cite{JLS} showed that in approximately
one million years of heavy tail ``Garch-trading'', with a reset every century,
{\it never} did three crashes similar to the three largest observed in the
true Dow Jones Index occur in a single ``Garch-century''. Of course, these
simulations
do not prove that our model is the correct one, only that one of the standard
models of the ``industry'' (which makes a reasonable null hypothesis) is
utterly unable to account for the stylized facts associated to large financial
crashes. What it suggests is that different mechanisms are responsible for
large crashes and that hypothesis 2 is the correct description of crashes.

\begin{figure}[b]
\begin{center}
\epsfig{file=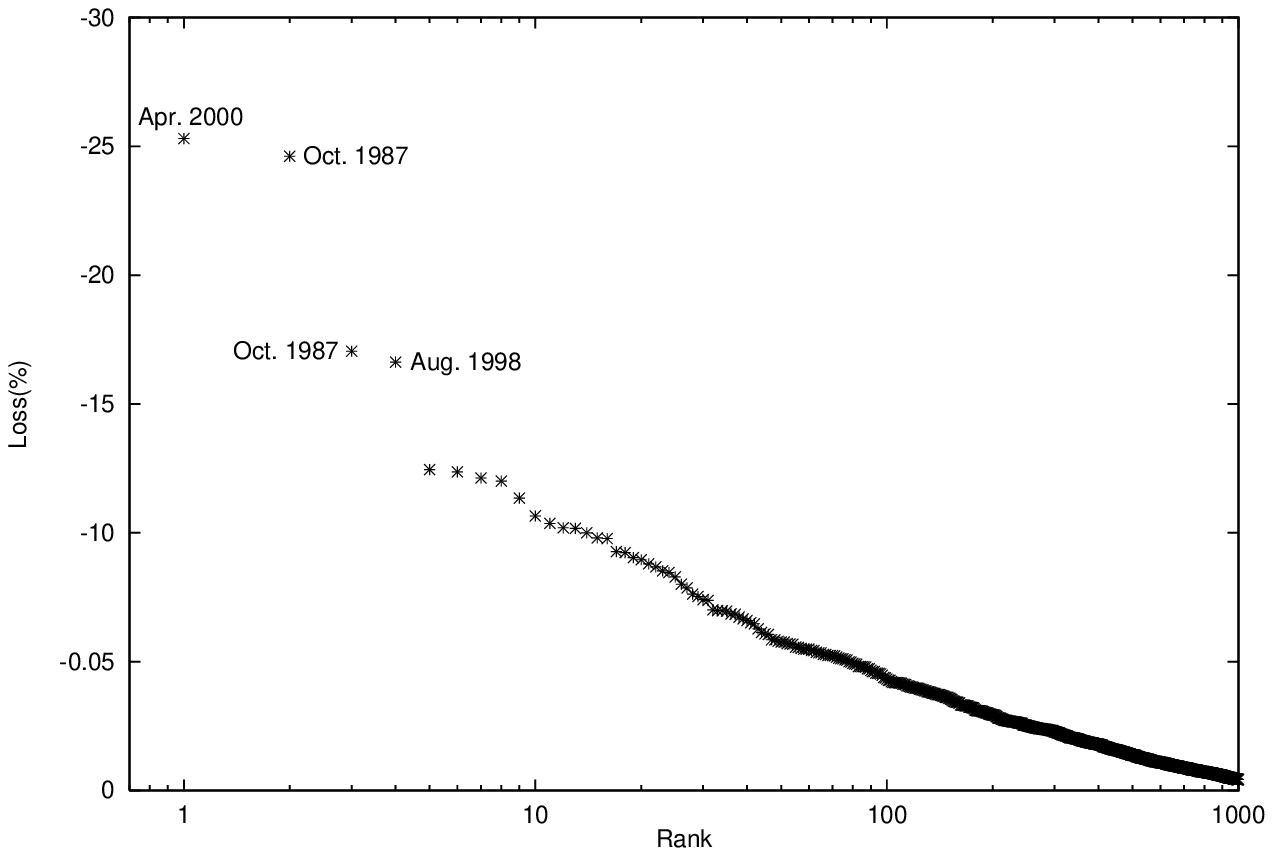}
\caption{\protect\label{nascumu}Rank ordering of draw downs
in the Nasdaq Composite since its establishment in 1971 until 18 April 2000.}

\vspace{5mm}

\epsfig{file=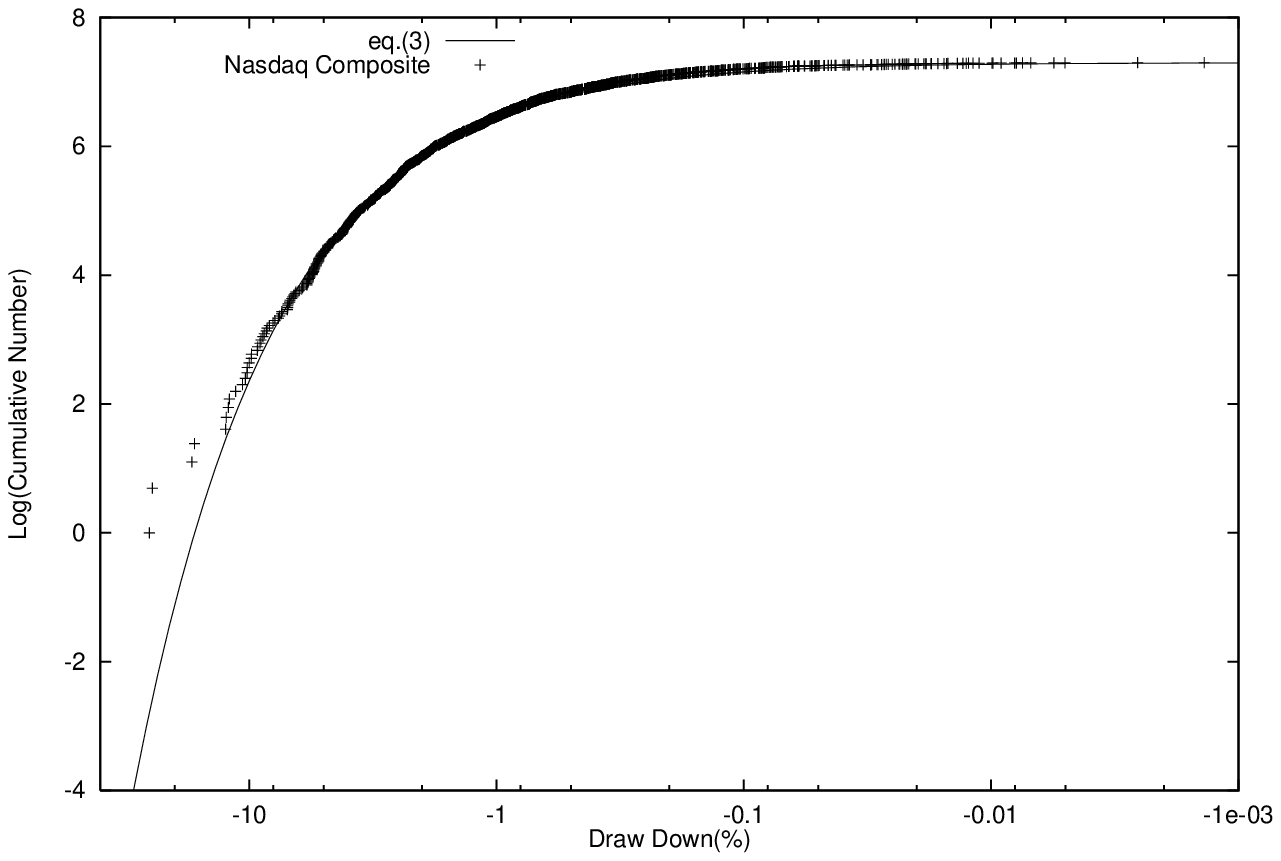}
\caption{\protect\label{nascumu1} Natural logarithm of the cumulative
distribution of draw downs $N(x)$ in the Nasdaq Composite since its
establishment in 1971 until 18 April 2000. The fit is $\ln(N) = \ln(1479)-29.0
x^{0.77}$ assuming that the distribution follows a stretched exponential
$N(x)=a\exp(-bx^c)$. Here $a=1479$ is the total number of draw downs. The
exponent $c \approx 0.8$ is compatible with values previously found in other
markets \protect\cite{lahsor,portsimo}.}
\end{center}
\end{figure}

A similar picture has been found for the Nasdaq Composite. In figure
\ref{nascumu}, we see the rank ordering plot of ``pure'' draw downs,
{\it i.e.}, no threshold (see \cite{JS98.1} for a brief discussion of the
effect of thresholds), since the establishment of the index in 1971 until
18 April 2000. Recall that the rank ordering plot, which is the same as the
(complementary) cumulative distribution with axis interchanged, puts
emphasis on
the largest events. Again, we see that the four largest events are not situated
on a continuation of the distribution of smaller events: the jump between
rank 4 and 5 in magnitude is $> 33\%$ whereas the corresponding jump between
rank 5 and 6 is $<1\%$ and this remains true for higher ranks. This means
that, for draw downs less than $12.5\%$, we have a more or less ``smooth''
curve
and then a $> 33\%$ gap(!) to rank 3 and 4. The four events are according
to rank the crash of April 2000 analysed here, the crash of Oct. 1987, a
$>17\%$ ``after-shock'' related to the crash of Oct. 1987 and a $>16\%$
drop related to the ``slow crash'' of Aug. 1998.

\begin{figure}[b]
\begin{center}
\epsfig{file=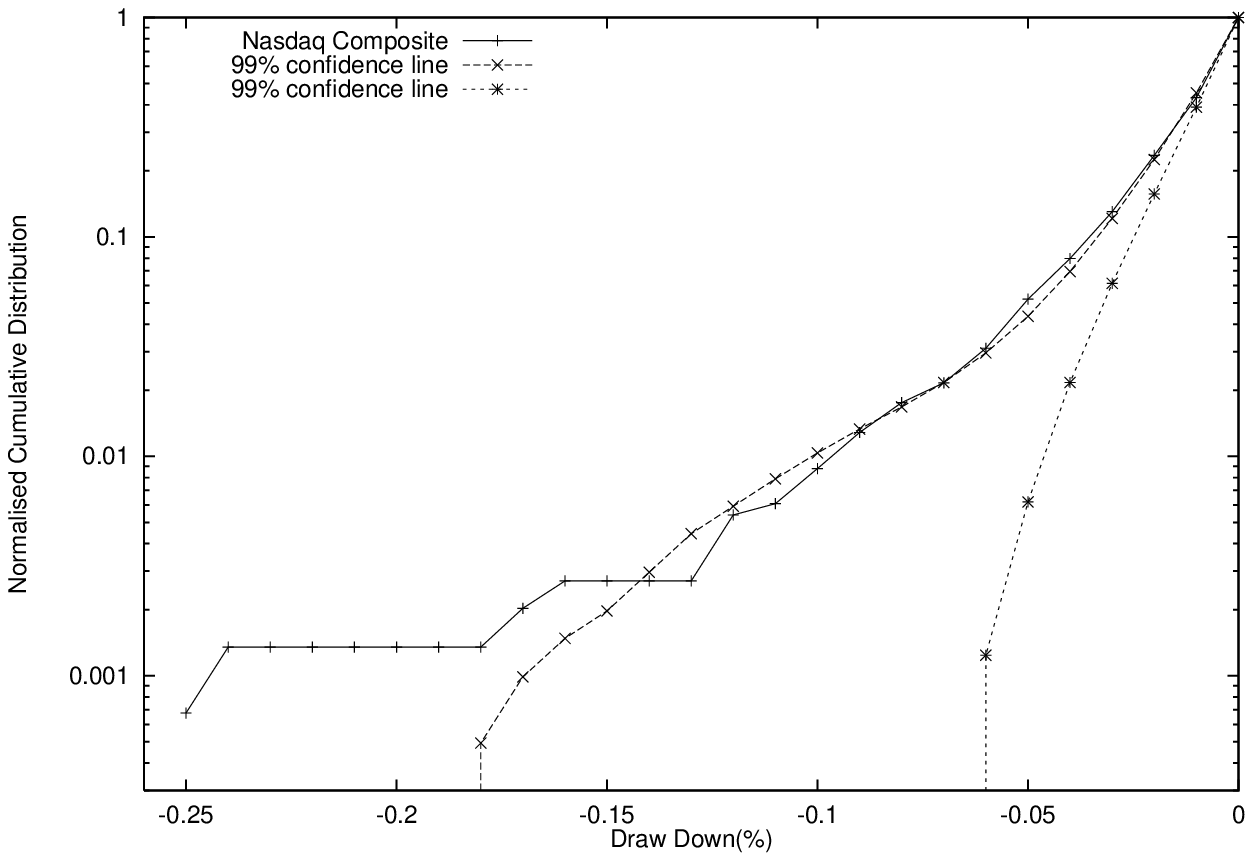}
\caption{\protect\label{confidence}Normalised cumulative distribution of draw
downs in the Nasdaq Composite since its establishment in 1971 until 18 April
2000. The 99\% confidence lines are estimated from the synthetic tests
described in section \protect\ref{out}.}
\vspace{5mm}

\epsfig{file=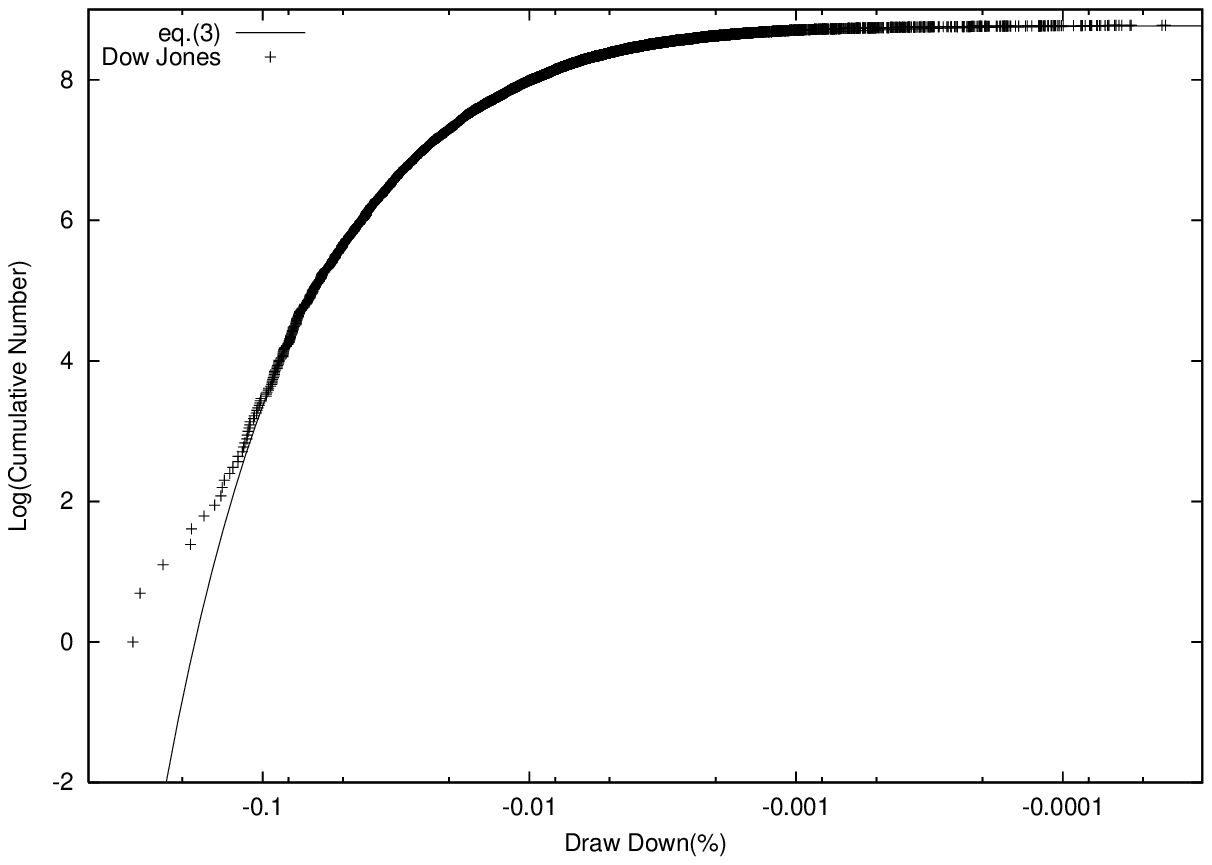}
\caption{\protect\label{djcumu1} Natural logarithm of the cumulative
distribution of draw downs $N(x)$ in the Dow Jones since 1900 until
2 May 2000. The fit is $\ln(N) = \ln(6469)-36.3 x^{0.83}$ assuming that
the distribution follows a stretched exponential $N(x)=a\exp(-bx^c)$.
Here $a=6469$ is the total number of draw downs. The exponent $c \approx 0.8$
is in remarkable agreement with the value found in the Nasdaq Composite,
see caption of figure \protect\ref{nascumu1}. The outliers to the fit are
according to rank the crash of Oct. 1987, the crash in 1914 related to the
outbreak of the First World War, the crash of Oct. 1929, two $>18\%$ crashes
in 1932 and 1933 respectively and two $>15\%$ "aftershocks" related to the
Oct 1929 crash.}
\end{center}
\end{figure}

In order to quantify the cumulative distribution of draw downs $N(x)$,
we compare it with a stretched exponential
\be
N(x) \approx a\exp(-bx^c)    \label{eq3}
\ee
as null-hypothesis \cite{lahsor}, see
figure \ref{nascumu1} and caption. Confirming the result
from the rank ordering, we see that the stretched exponential captures well
the distribution except for the four largest events. Furthermore, it is clear
from figure \ref{nascumu1} that the distribution is not that of a power law
which would be qualified as a straight line in this log-log plot. One could
perhaps argue
that the tail of the cumulative distribution tends to become linear in this
log-log
representation; however, this observation is based on an interval smaller than
half-a-decade. In addition, in the case of the Dow Jones data shown in
figure \ref{djcumu1},
the tail is even fatter than a power law with an upward convexity in the
log-log
representation. The principle
of parsimony leads us to prefer not to assume any distribution in the tail and
only conclude about the fact that the few largest events are clearly taken
from a different distribution. Indeed, if we
extrapolate the curve to larger events, we get that, in the $\approx 30$ years
of the existence of the Nasdaq Composite, we should have observed $0.09$
draw downs above $24\%$, whereas in reality we have observed $2$.

To further establish the statistical confidence with which we can conclude
that the four largest events are outliers, we have reshuffled the daily
returns 1000 times and hence generated 1000 synthetic data sets. This
procedure means that the synthetic data will have exactly the same
distribution of daily returns. However, higher order correlations apparently
present in the largest draw downs are destroyed by the reshuffling. This
surrogate data analysis of the distribution of draw downs has the advantage
of being {\it non-parametric}, {\it i.e.}, independent of the quality of fits
with a model such as the stretched exponential or the power law. We will
now compare the
distribution of draw downs both for the real data and the synthetic data.
With respect to the synthetic data, this can be done in two complementary ways.
In figure \ref{confidence}, we see the distribution of draw downs in the
Nasdaq Composite compared with the two lines constructed at the $99\%$
confidence level
for the entire
{\it ensemble} of synthetic draw downs, {\it i.e.} by considering the
individual draw downs
as independent: for any given draw down, the upper (resp. lower)
confidence line is such that $5$ of the synthetic distributions are above
(below) it;
as a consequence, 990 synthetic times series out of the 1000 are within the
two confidence lines for any draw down value which define the typical interval
within which we expect to find the empirical distribution.

Two features are apparent on figure \ref{confidence}. First,
the distribution of the true data
breaks away from the $99\%$ confidence intervals at $\approx 15\%$, showing
that the four largest events are indeed outliers in this sense.
In addition, the empirical distribution of draw downs is systematically
found close to the
upper confidence boundary, with an upward curvature described by the
apparent stretched exponential eq. (\ref{eq3}), for values
less than $15\%$. In contrast,
the median value between the two confidence lines
is approximately linear in this semi-logarithmic representation, qualifying an
exponential distribution as expected for uncorrelated daily returns (see
appendix 1 of
\cite{JLS}). The upward curvature of the distribution of draw downs and
its closeness to the upper confidence line
thus signals a subtle dependence between consecutive returns.

A more sophisticated analysis is to consider each synthetic data set
{\it separately} and calculate the {\it conditional probability} of
observing a given draw down given some prior observation of draw downs.
This gives a more precise estimation of the statistical significance
of the outliers, because the previously defined confidence lines neglect
the correlations created by the ordering process
which is explicit in the construction of a cumulative distribution.

Out of the 10.000 synthetic data sets, 776 had a single draw down larger
than $16.5\%$, 13 had two draw downs larger than $16.5\%$, 1 had three draw
downs larger than $16.5\%$ and none had 4 (or more) draw downs larger than
$16.5\%$ as in the real data. This means that given the distribution of
returns, by chance we have a $\approx 8\%$ probability of observing a draw
downs larger than $16.5\%$, a $\approx 0.1\%$ probability of observing two
draw downs larger than $16.5\%$ and for all practical purposes zero
probability of observing three or more draw downs larger than $16.5\%$.
Hence, the probability that the largest four draw downs observed for the
Nasdaq could result from chance is outside a $99.99\%$ confidence interval.
As a consequence we are lead to conclude that the largest market events are
to be characterised by the presence of higher order correlations in contrast to
what is observed during ``normal'' times.

Performing a fit with a stretched exponential on the cumulative distribution
of ``pure'' draw downs in the Dow Jones index since 1900 until May 2000,
{\it i.e.}, no threshold as for the Nasdaq Composite, gives a
remarkably similar result as that of the Nasdaq Composite, see figure
\ref{djcumu1} and caption, both with respect to the exponent as well as the
point were the data ``breaks away'' from the fit. If we again extrapolate the
fit to the largest events, we get that $0.12$ draw downs above $23.5\%$
should have occured in the last century whereas we in fact have observed $3$.

This analysis confirms the conclusion from the previous analysis of the Dow
Jones that draw downs larger than $\approx 15$\% are to be considered as
outliers with high probability. It is interesting that the same amplitude
of $\approx 15$\% is found for both markets considering the much larger
daily volatility of the Nasdaq Composite. This may result from the fact that,
as we have shown, very large draw downs are more controlled by transient
correlations than by the amplitude of daily returns.

The presented statistical analysis of the Dow Jones Average and the Nasdaq
Composite suggests that large crashes {\it are} special and that precursory
patterns may exist decorating the speculative bubble preceding the crash.
As we have discussed in the preceding section, such precursory patterns do
exist prior to these outliers and can be quantified by eq. (\ref{lpeq}). It is
the subject of the next section to quantify these precursory patterns for the
Nasdaq bubble starting in spring 1997 and ending April 2000 and to compare
with the results previously obtained for large crashes.

\section{The current crash}

With the low of 3227 on the 17 April, the Nasdaq Composite lost over
$37 \%$ of its all-time high of 5133 reached on the 10'th of March this year.
The Nasdaq Composite consists mainly of stock related to the so-called
``New Economy'', {\it i.e.}, the Internet, software, computer hardware,
telecommunication ... A main characteristic of these companies is that their
price-earning-ratios (P/E's), and even more so their price-dividend-ratios,
often come in three digits\footnote{VA LINUX to be discussed below actually
has a {\it negative} Earning/Share of -1.68. Yet they are currently traded
around \$40 per share which is close to the price of Ford in early March
2000.}. Opposed to this, so-called ``Old
Economy'' companies, such as Ford, General Motors and DaimlerChrysler, have
P/E $\approx 10$. The
difference between ``Old Economy'' and ``New Economy'' stocks is
thus the expectation of {\it future earnings} as discussed in \cite{ds}:
investors expect an enormous increase in for example the sale of Internet and
computer related products rather than in car sales and are hence more
willing to invest in Cisco rather than in Ford notwithstanding the fact
that the earning-per-share of the former is much smaller than for the
later. For a similar price per share (approximately \$60 for Cisco and \$55
for Ford), the earning per share is \$0.37 for Cisco compared to
\$6.0 for Ford (Cisco has a total market capitalisation of \$395 billions
(close of April, 14, 2000) compared to \$63 billions for Ford).
In the standard fundamental valuation formula, in which the expected return of
a company is the sum of the dividend return and of the growth rate, ``New
Economy'' companies are supposed to compensate for their lack of present
earnings
by a fantastic potential growth. In essence, this means that the bull market
observed in the Nasdaq the last three years until recently is fueled by
expectations of increasing future earnings rather than economic
fundamentals: the
price-to-dividend ratio for a company such as Lucent Technologies (LU) with a
capitalization of over $\$300$ billions prior to its crash on the 5 Jan.
2000 is
over $900$ which means that you get a higher return on your checking
account(!)
unless the price of the stock increases. Opposed to this, an ``Old
Economy'' company
such as DaimlerChrysler gives a return which is more than thirty times higher.
Nevertheless, the shares of Lucent Technologies rose by more than $40\%$
during 1999 whereas the share of DaimlerChrysler declined by more than $40$\%
in the same period.
Truly surrealistic is the fact that the recent crashes of IBM, LU and
Procter \& Gamble (P\&G) correspond to a loss equivalent to many countries
state budget! And this is usually attributed to a ``business-as-usual''
corporate statement of a slightly revised smaller-than-expected earnings!

These considerations makes it clear that it is the {\it expectation} of
future earnings that motivates the average investor rather than present
economic
reality, thus creating a speculative bubble. History provides many examples
of bubbles driven by unrealistic expectations of future earnings followed by
crashes \cite{white}. The same basic ingredients are found repeatedly:
fueled by initially well-founded economic fundamentals, investors
develop a self-fulfilling enthusiasm by an imitative process or crowd
behavior that leads to the building of ``castles in the air'', to paraphrase
Malkiel \cite{Malkiel}. Furthermore, the causes of the crashes on
the U.S. markets in 1929, 1987, 1998 and present belongs to the same category,
the difference being mainly in which sector the bubble was created: in 1929, it
was utilities; in 1987, the bubble was supported by a general deregulation of
the market with many new private investors entering the market with very high
expectations with respect to the profit they would make; in 1998, it was an
enormous expectation
to the investment opportunities in Russia that collapsed; as for the
present, it is
the extremely high expectations to the Internet, telecommunication etc.
that has
fueled the bubble. The IPO's (initial public offerings)
of many Internet and software companies has been
followed by a mad frenzy where the price of the share has soared during the
first few hours of trading. An excellent example is VA LINUX SYSTEMS whose
\$$30$ IPO price increased a record 697 percent to close at \$$239.25$ on its
opening day 9 Dec. 1999, only to decline to \$$28.94$ on the 14 April 2000.

\begin{figure}
\begin{center}
\epsfig{file=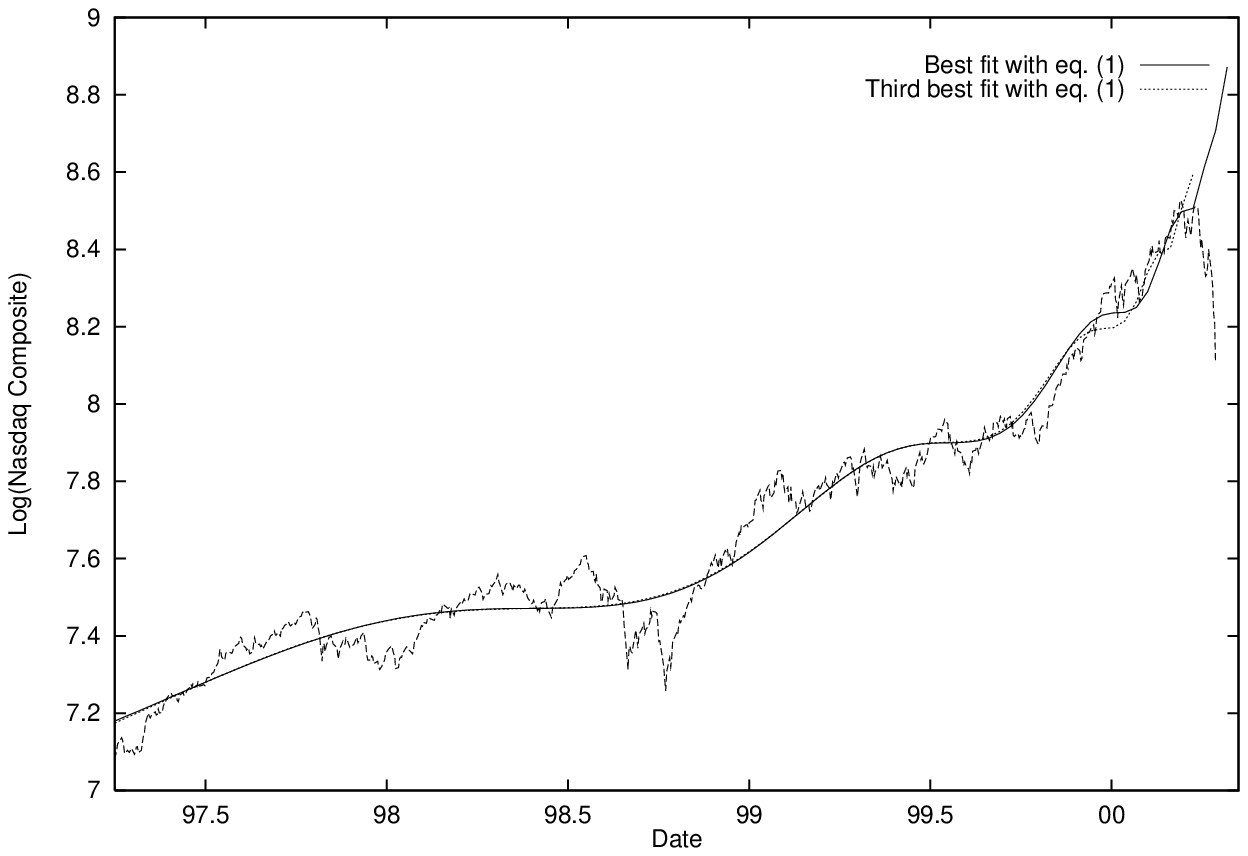}
\caption{\protect\label{nasfit} Best (r.m.s. $\approx 0.061$) and third best
(r.m.s. $\approx 0.063$) fits with eq. (\protect\ref{lpeq}) to the natural
logarithm of the Nasdaq Composite. The parameter
values of the fits are $A\approx 9.5$, $B\approx -1.7$, $C\approx 0.06$,
$z\approx 0.27$, $t_c\approx 2000.33$, $\omega \approx 7.0$, $\phi \approx
-0.1$ and $A\approx 8.8$, $B\approx -1.1$, $C\approx 0.06$ ,$z\approx 0.39$,
$t_c\approx 2000.25$,$\omega \approx 6.5$, $\phi \approx -0.8$, respectively.}

\vspace{5mm}

\epsfig{file=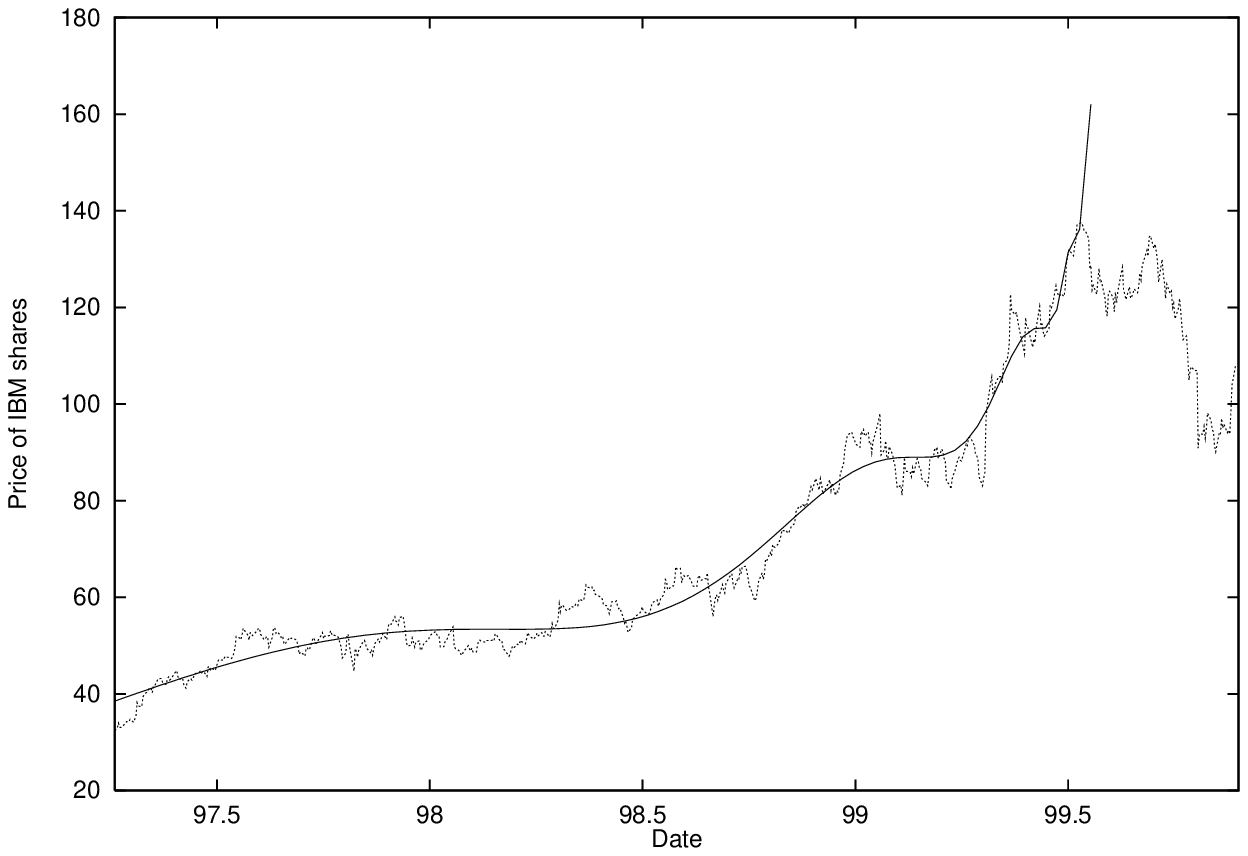}
\caption{\protect\label{ibm} Best (r.m.s. $\approx 3.7$) fit with eq.
(\protect\ref{lpeq}) to the price of IBM shares. The parameter
values of the fits are $A\approx 196$, $B\approx -132$, $C\approx -6.1$,
$z\approx 0.24$, $t_c\approx 99.56$, $\omega \approx 5.2$ and
$\phi \approx 0.1$}
\end{center}
\end{figure}

In figure \ref{nasfit}, we see the logarithm of the Nasdaq Composite fitted
with eq. (\ref{lpeq}). The data interval to fit was identified using the same
procedure as for the other crashes: the first point is the lowest value of
the index prior to the onset of the bubble and the last point is that of the
all-time high of the index. There exists some subtelty with respect to
identifying the onset of the bubble, the end of the bubble being objectively
defined as the date where the market reached is maximum. A bubble signifies
an acceleration of the price. In the case of Nasdaq, it tripled from 1990 to
1997. However, the increase was a about factor 4 in the 3 years preceding
the current crash thus defining an ``inflection point'' in the index. In
general,
the identification of such an ``inflection point'' is quite straightforward on
the most liquid markets whereas this is not the case for the emergent markets.
With respect to details of the methodology of the fitting procedure, we refer
the reader to \cite{JSL,Thesis}.

Three fits were obtained with similar parameter values for the best and
third best
fit, whereas the second best fit had a rather small value for $z \approx
0.08$ and
a rather high value for $\omega \approx 7.9$ compared with previous results
and
is not shown. The values obtained for the best and third best fit are $\omega
\approx 7.0$ and $\omega \approx 6.5$, $z\approx 0.27$ and $z\approx 0.39$,
$t_c \approx 2000.34$ and $t_c \approx 2000.25$, respectively. These results
pointed to a crash occuring between the 31 of March 2000 and 2 May 2000 and
have now been confirmed by the recent market event.

\section{Prediction}

An obvious question concerns the predictive power of eq. (\ref{lpeq}). In the
present case, the last point used in the fitted data interval was that of
March 10, 2000. The predicted time of the crash was as mentioned 2 May for the
best fit and and 31 March for the third best fit. Except for slight gains on
31 March and 5, 6 and 7 April, the closing of the Nasdaq Composite
has been in continuous decline since the 24 March and lost over $25$ \% in the
week ending on Friday the 14 April. Consequently, the crash occured
approximately in between the predicted date of the two fits. The corresponding
dates for the 1929, 1987 and 1998 crashes on Wall Street and the 1987, 1994 and
1997 crashes on the Hong-Kong stock exchange as well as the collapse of the
US\$ in 1985 are shown in table \ref{table1}
for comparison.
\begin{table}[h]
\begin{center}
\begin{tabular}{|c|c|c|c|c|c|c|c|c|c|} \hline
crash & $t_c$ & $t_{max}$ & $t_{min}$ & $\%$ drop & $z$ & $\omega$ &
$\lambda$ \\ \hline
1929 (DJ) &  $30.22$ & $29.65$ & $29.87$ & $47\%$ & $0.45$ & $7.9$ & $2.2$\\
\hline
1985 (DM) &  $85.20$ & $85.15$ & $85.30$ & $14\%$ & $0.28$ & $6.0$ & $2.8$
\\ \hline
1985 (CHF) &  $85.19$ & $85.18$ & $85.30$ & $15\%$ & $0.36$ & $5.2$ & $3.4$
\\ \hline
1987 (S\&P)&  $87.74$ & $87.65$ & $87.80$ & $30\%$ & $0.33$ & $7.4$ & $2.3$\\
\hline
1987 (H-K)&  $87.84$ & $87.75$ & $87.85$ & $50\%$ & $0.29$ & $5.6$ & $3.1$\\
\hline
1994 (H-K)&  $94.02$ & $94.01$ & $94.04$ & $17\%$ & $0.12$ & $6.3$ & $2.7$ \\
\hline
1997 (H-K) &  $97.74$ & $97.60$ & $97.82$ & $42\%$ & $0.34$ & $7.5$ & $2.3$ \\
\hline
1998 (S\&P) &  $98.72$ & $98.55$ & $98.67$ & $19.4\%$ & $0.60$ & $6.4$ &
$2.7$\\
\hline
1999 (IBM) & $99.56$ &$99.53$ & $99.81$  & $34\%$ & $0.24$ & $5.2$ & $3.4$\\
\hline
2000 (P\&G) & $00.04$ & $00.04$ & $00.19$ & $54\%$ & $0.35$ & $6.6$ & $2.6$\\
\hline
2000 (Nasdaq) & $00.34$ & $00.22$ & $00.29$ & $37\%$ & $0.27$ & $7.0$ & $2.4$\\
\hline
\end{tabular}
\end{center}
\caption{\label{table1} $t_c$ is the critical time predicted from the fit
of the
financial time series to the eq. (\ref{lpeq}). The other parameters $z$,
$\omega$ and $\lambda$ of the fit are also shown. The fit is performed up to
the time $t_{max}$ at which the market index achieved its highest maximum
before the crash. $t_{min}$ is the time of the lowest point of the market
before rebound. The percentage drop is calculated from the total loss from
$t_{max}$ to $t_{min}$.
}
\end{table}

We see that, in all 9 cases, the market crash started at a time between the
date of the last point and the predicted $t_c$. And with the exception of the
Oct. 1929 crash and using the third best fit of the present crash
(this fit had $\omega/2\pi \approx 1$) in all cases the market ended its
decline less than approximately one month after the predicted $t_c$. These
results indeed suggest that predictions of crashes with eq. (\ref{lpeq}) is
indeed possible.

\begin{figure}
\begin{center}
\epsfig{file=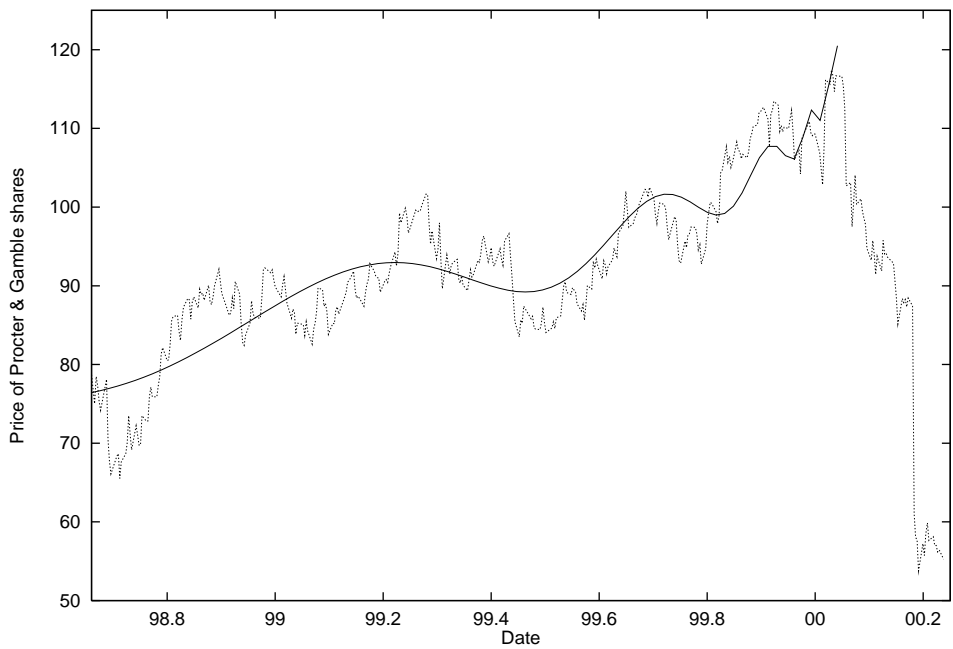}
\caption{\protect\label{pg} Best (r.m.s. $\approx 4.3$) fit with eq.
(\protect\ref{lpeq}) to
the price of Procter \& Gamble shares. The parameter values of the fit are
$A\approx 124$, $B\approx -38$, $C\approx 4.8$ ,$z\approx 0.35$, $t_c\approx
2000.04$, $\omega \approx 6.6$ and $\phi \approx -0.9$.}
\end{center}
\end{figure}

Furthermore, the crashes of the shares of IBM, LU and P\&G, {\it i.e.},
three of the largest U.S. companies, may be taken as precursors of a pending
crash signifying how unstable the market actually was in the months preceding
the current crash. Quite remarkably, two of these three company crashes were
also preceded by a speculative bubble with the same characteristics as
previously seen on the market as a whole, see figures \ref{pg}, \ref{ibm}
and table \ref{table1}.

Of course, the results presented here does not mean that we have publicly
predicted the April 2000 crash of the Nasdaq Composite. This has neither
been the purpose. What the analysis presented above shows is that eq.
(\ref{lpeq}) has predictive power. Furthermore, from a purely scientific point
of view, it is the {\it observation} and the {\it comparison} between the
observation and the predictions of the model which carry meaning.

\section{False alarms}

Not all speculative bubbles end in a crash. Hence, the question about false
alarms enters naturally. We have twice identified a log-periodic power law
bubble signaling a crash where the market in fact {\it did not} crash
according to the definition presented in section \ref{out}.
The first attempt was in Oct. 1997 where the market dropped only $7$\%
\cite{oct97} and quickly recovered (see also \cite{Van2}). The second
attempt was in October last year when the world markets were sent into
turmoil by a speech by Alan Greenspan and the Dow Jones for the first time
since 8 April 1999 dipped below 10.000 on the 15 and 18 Oct 1999. However,
the market did
not crash and instead quickly recovered. These two examples of bubbles
landing more or less smoothly are completely consistent with the theory of
rational bubbles and crashes developed in \cite{JSL}. This also illustrates
the difficulties involved in a crash-prediction scheme using eq. (\ref{lpeq}):
according to the theory, the critical time $t_c$ is not necessarily the time
of the crash, only its most probable time; in addition, there is a finite
probability that the bubble ends without crashing. We are
currently investigating how to extend the methodology in order to increase
the reliability of the model in terms of predictions.

\section{Conclusion}

Here, we have provided yet an example of a speculative bubble with
power law acceleration and log-periodic oscillations ending in a crash/major
correction, {\it i.e.,} that of the Nasdaq Composite starting in spring 1997
ending
in late March/early April 2000. The log-frequency of these oscillations
is in remarkable agreement with what has been obtained previously on a wide
range of markets \cite{JSL,Emerg}.

The present analysis of these market phases emphasizes a collective
behavior of
investors, leading to a fundamental ripening of the markets towards an
instability.
This must be contrasted with
the endeavor of economists and analysts who search for contemporary
news to explain the events. For instance, the Aug. 1998 crash was often
attributed
to a devaluation of the ruble and to events on the Russian political scene.
While
we do not underestimate the effect of ``news'', we observe that markets are
constantly
``bombarded'' by news and it will always be possible to attribute the crash to
a specific one, {\it after the fact}. In contrast, we
view their reactions more often than not as reflecting their
underlying stability (or instability). In the case of the Aug. 1998 crash,
the market was ripe for
a correction and the ``news'' made it occur. If nothing had occurred on the
Russian
scene, we have proposed \cite{JSL} that other news would have
triggered the event anyway, within a time scale of about a month, which
seems to be the relevant lifetime of a
market instability associated with the burst of a bubble. With respect to the
present Nasdaq crash, undoubtedly, analysts will forge post-mortem stories
linking it in part with the  effect of the crash of Microsoft Inc. resulting
from the breaking of negotiations during the week-end of April 1st
with the US federal government on the antitrust issue. Again, we see
the Nasdaq crash as the natural death of a speculative bubble, anti-trust
or not, the results presented here strongly suggest that the bubble would
have collapsed anyway.

However, according to our analysis (see the probabilistic model of bubbles in
\cite{risk,JSL,JLS}),
the exact timing of its death is not fully deterministic and allows
for stochastic influences, but within the remarkably tight bound of about
one month.

We have also discussed the possibility of using the proposed framework,
specifically eq. (\ref{lpeq}), in order to predict when the market will
exhibit a crash/major correction. Our analysis not only points to a
predictive potential but also that false alarms are difficult to avoid
due to the underlying nature of speculative bubbles.

A fundamental remaining question concerns the use of a reliable crash
prediction scheme. Assume that a crash prediction is issued stating that a
crash will occur $x$ weeks from now. At least three different scenarios
are possible:
\begin{itemize}

\item Nobody believes the prediction which was then futile and, assuming
that the prediction was correct, the market crashes\footnote{One may consider
this as a victory for the ``predictors'' but as we have experienced in relation
to our quantitative prediction of the change in regime of the Nikkei index
\protect\cite{nikepred}, this would only be considered by some
critics just another
``lucky one'' without any statistical significance (see \cite{japassess} for
an alternative Bayesian approach).}.

\item Everybody believes the warning, which causes panic and the market
crashes as consequence. The prediction hence seems self-fulfilling.

\item Enough believe that the prediction {\it may} be correct and the steam
goes off the bubble. The prediction hence disproves itself.

\end{itemize}

None of these scenarios are attractive. In the first two, the crash is not
avoided and in the last scenario the prediction disproves itself and as
a consequence the theory looks unreliable. This seems to be the unescapable
lot of scientific investigations of systems with learning and reflective
abilities, in contrast with the usual inanimate and unchanging physical laws
of nature. Furthermore, this touches the key-problem of scientific
responsibility. Naturally, scientists have a responsibility to publish their
findings. However, when it comes to the practical implementation of those
findings in society, the question becomes considerably more complex.

\vskip 0.5cm
{\bf Acknowledgements}: We thank D. Stauffer for stimulating comments on
the manuscript. D.S. also thanks B. Roehner for his comments.

\newpage

\end{document}